\documentclass[floats,twocolumn,aps,prb,longbibliography,superscriptaddress]{revtex4-2}

\usepackage{graphicx}
\usepackage{amsfonts}
\usepackage{amssymb}
\usepackage{amsmath}
\usepackage{dsfont}
\usepackage{amsbsy}
\usepackage[colorlinks=true, urlcolor=magenta, linkcolor=magenta, citecolor=magenta, pdftex]{hyperref}
\usepackage{hyperref}
\usepackage{xcolor}
\usepackage{subeqnarray}
\usepackage{tabularx}
\usepackage{tabularray}

\begin{document}

\title{Reentrant behavior and possible $2/3$ magnetization plateau on the double-trillium langbeinite K$_2$Ni$_2$(SO$_4$)$_3$}

\author{Mat\'ias G. Gonzalez}
\affiliation{Institute of Physics, University of Bonn, Nussallee 12, 53115 Bonn, Germany}

\author{Yurii Skourski}
\affiliation{Dresden High Magnetic Field Laboratory (HLD-EMFL), Helmholtz-Zentrum Dresden-Rossendorf, 01328 Dresden, Germany\looseness=-1}

\author{Johannes Reuther}
\affiliation{Helmholtz-Zentrum Berlin f\"ur Materialien und Energie, Hahn-Meitner-Platz 1, 14109 Berlin, Germany}
\affiliation{Dahlem Center for Complex Quantum Systems and Fachbereich Physik, Freie Universität Berlin, 14195 Berlin, Germany\looseness=-1}

\author{Ivica {\v Z}ivkovi{\' c}}
\affiliation{Laboratory for Quantum Magnetism, Institute of Physics, \'Ecole Polytechnique F\'ed\'erale de Lausanne, CH-1015 Lausanne, Switzerland\looseness=-1}

\begin{abstract} 
K$_2$Ni$_2$(SO$_4$)$_3$ is a member of the langbeinite family, consisting of two intertwined $S=1$ trillium lattices, out of which one is strongly coupled (strong-TL) and the other is weakly coupled (weak-TL). Further inter-trillium interactions give rise to a highly-frustrated Heisenberg Hamiltonian. Despite ordering at low temperatures, K$_2$Ni$_2$(SO$_4$)$_3$ lies close in parameter space to a spin-liquid region that surrounds the tetratrillium limit, where each triangle belonging to strong-TL turns into a tetrahedron by connecting to a single spin from weak-TL. Here, we compare the experimentally determined magnetization process using pulsed magnetic fields up to $40$ T with classical Monte Carlo calculations, uncovering a series of phase transitions at both low and intermediate fields. Furthermore, we reveal a signature of a $2/3$ magnetization plateau consisting of a $1/3$ phase on strong-TL and a fully polarized phase on weak-TL. Although in the classical limit no plateau is expected, we find a very prominent dome structure reflecting the tendency of the system to stabilize this particular spin configuration. The presence of a dome leads to a reentrant phenomenon in which the system recovers the Hamiltonian symmetries when increasing the magnetic field. Finally, we show that this plateau-like phase is also present in the classical Heisenberg model on the single trillium and tetratrillium lattices, indicating its possible presence in the large family of double-trillium langbeinite compounds. Our findings motivate future studies on the presence of the plateau phase in the quantum limit of both trillium and double-trillium materials within the langbeinite family.
\end{abstract}

\date{\today}

\maketitle

\section{Introduction}

The term \textit{frustrated magnetism} refers to spin systems in which competing interactions prevent the simultaneous minimization of all pairwise interaction energies in the ground state. Several mechanisms can give rise to competing interactions in magnetic systems; among these, geometric frustration is the primary source in antiferromagnetic Heisenberg (i.e., spin-isotropic) models. The simplest examples of magnetic frustration arise in non-bipartite lattices when nearest-neighbor interactions are sufficient to create competing interactions, for example, in the triangular and kagome lattices. In other cases, frustration can be increased by including further neighbor interactions. This is especially relevant in three-dimensional (3D) materials, where the number of superexchange paths between magnetic ions is generally larger, and the distances between neighboring sites are more similar than in two-dimensional systems. Consequently, it is not surprising that a large number of highly frustrated 3D materials have been identified in recent years, particularly in the search for spin-liquid candidates. Some examples are the hyper-hyperkagome material PbCuTe$_2$O$_6$~\cite{Chillal20, Chern21, Fancelli23}, the orthovanadate HoVO$_4$~\cite{Ranaut22}, and the body-centered cubic garnet NaCa$_2$Cu$_2$(VO$_4$)$_3$~\cite{Alexanian25}.

In this context, also the langbeinite family of compounds has gained considerable attention in recent years~\cite{Zivkovic21, Yao23, Gonzalez24, Boya22, Sebastian25, Khatua25, Kubickova24, Khatua24, Kolay24, Magar26, Khatua25b}. Langbeinites have two symmetry-inequivalent magnetic sites forming two intertwined trillium lattices. One trillium lattice alone is already frustrated, consisting of tri-coordinated triangles, which means that each spin belongs to three distinct triangles. Even though frustrated, the spins order in the classical limit at $T_c\simeq 0.21~J$ forming a single-$\mathbf{Q}$ state~\cite{Hopkinson06, Isakov08}. The double trillium lattice structure of the langbeinite family gives rise to an extra source of frustration from inter-trillium couplings, providing a promising framework in the search for spin liquid candidates. So far, several langbeinites have been synthesized and studied such as the $S=1$ K$_2$Ni$_2$(SO$_4$)$_3$~\cite{Zivkovic21, Yao23, Gonzalez24}, the $S=2$ Cs$_2$Fe$_2$(MoO$_4$)$_3$~\cite{Kubickova24}, the $S=5/2$ KSrFe$_2$(PO$_4$)$_3$~\cite{Boya22, Sebastian25} and K$_2$FeSn(PO$_4$)$_3$~\cite{Khatua25}, the $S=3/2$ K$_2$CrTi(PO$_4$)$_3$~\cite{Khatua24} and KBaCr$_2$(PO$_4$)$_3$~\cite{Kolay24}, and possible $S=1/2$ realizations in K$_2$Co$_2$(SO$_4$)$_3$~\cite{Magar26}, among others. These materials exhibit different properties depending on the interplay between the five most short-range couplings, two of which define trillium lattices, while the remaining three connect them. For example, KSrFe$_2$(PO$_4$)$_3$ is a spin liquid candidate which displays no sign of a phase transition down to low temperatures~\cite{Boya22}. Others like KBaCr$_2$(PO$_4$)$_3$ have an ordering phase transition attributed to ferromagnetic couplings~\cite{Kolay24}. Measurements on K$_2$CrTi(PO$_4$)$_3$ reveal two phase transitions at different temperatures~\cite{Khatua24}. Altogether, these differing results indicate that the langbeinite family of materials exhibits a complex and rich landscape of magnetic phases.

In this article, we focus on K$_2$Ni$_2$(SO$_4$)$_3$, where the Ni$^{2+}$ ions can be represented by $S=1$ spins. This compound has a Curie-Weiss temperature of $\theta_{\text CW} = -18$~K, and shows a magnetic phase transition at $T_N \sim 1.1$~K~\cite{Zivkovic21, Yao23, Gonzalez24} where only about $1\%$ of the entropy is released~\cite{Zivkovic21} and the spectrum of excitations is dominated by a broad continuum even at $T\sim 0.1~T_N$~\cite{Yao23}. These features reveal a highly dynamic state whose origin may be related to the proximity in parameter space to a phase that has been theoretically predicted to be a quantum spin liquid down to $T=0$~\cite{Gonzalez24}. This quantum spin liquid arises from a specific coupling pattern in which each site of the weakly-coupled trillium lattice (weak-TL) is connected to a triangular plaquette of the strongly-coupled lattice (strong-TL), together forming a tetrahedral unit. Therefore, this lattice has been referred to as the tetratrillium lattice and exhibits a spin liquid ground state both in the classical and quantum limits~\cite{Gonzalez24, Gonzalez26}.

In this article, we use both theoretical and experimental methods to study the behavior of K$_2$Ni$_2$(SO$_4$)$_3$ in a magnetic field. The magnetic field adds an additional source of frustration to the system, which drives the emergence of interesting gapped phases that live on magnetization plateaus~\cite{Takigawa11}. These plateaus are of purely quantum origin and have been studied in a wide variety of frustrated spin systems such as the triangular, kagome, and pyrochlore lattices~\cite{Honecker99, Penc04, Sakai11, Nakano17, Schnack18, Misawa20, Schulter22, Hagymasi22, Morita23, He24}. Despite the quantum origin of the plateaus, classical calculations can, in some cases, evidence their existence at finite temperatures, for example, in the triangular lattice~\cite{Seabra11}. 

To study the magnetization process of K$_2$Ni$_2$(SO$_4$)$_3$, we perform pulsed magnetic field measurements on single-crystal samples up to $40$~T to reach the saturation field, and thus obtain the full magnetization curve at several temperatures. Our measurements evidence a sharper increase in the magnetization for low magnetic fields, which then continues approximately linearly towards the saturation field at $\sim 25$~T. Calculating the field derivative $dM/dH$ reveals the presence of hidden details, with several features indicating possible transitions between field-induced phases. We compare these findings with classical Monte Carlo (cMC) calculations using the exchange couplings derived in Ref.~\cite{Gonzalez24}, which capture the ordering wave-vector at zero field correctly and reproduce the inelastic neutron scattering spectra at finite temperatures. Here, we find that these parameters also reproduce the saturation field accurately, providing further indication of the model's accuracy. Our calculations also evidence a complicated succession of transitions at small magnetic fields, in accordance with the already established sensitivity of the system to small deviations in exchange parameters~\cite{Gonzalez24}. At intermediate fields, the classical calculations reveal the existence of a 5-up-1-down phase which would correspond to a $2/3$ magnetization plateau in the quantum case. In this phase, weak-TL is fully polarized, while strong-TL presents a $1/3$ phase. This phase stands out prominently in the temperature-field phase diagram, exhibiting a dome and resulting in a reentrant behavior for a certain range of temperatures when increasing the magnetic field.

The remainder of the article is organized as follows. In Section~\ref{sec:mod}, we review established properties of K$_2$Ni$_2$(SO$_4$)$_3$, including its exchange Hamiltonian, and outline the experimental and theoretical methods employed in this work. In Section~\ref{sec:res}, we present and discuss our results, beginning with data from high pulsed-field measurements, followed by results from cMC calculations. Finally, in Sec.~\ref{sec:conc} we present the final discussion and conclusions of our work.

\section{Model and Methods}
\label{sec:mod}

\begin{figure}[t!]
    \centering
    \includegraphics[width=0.95\linewidth]{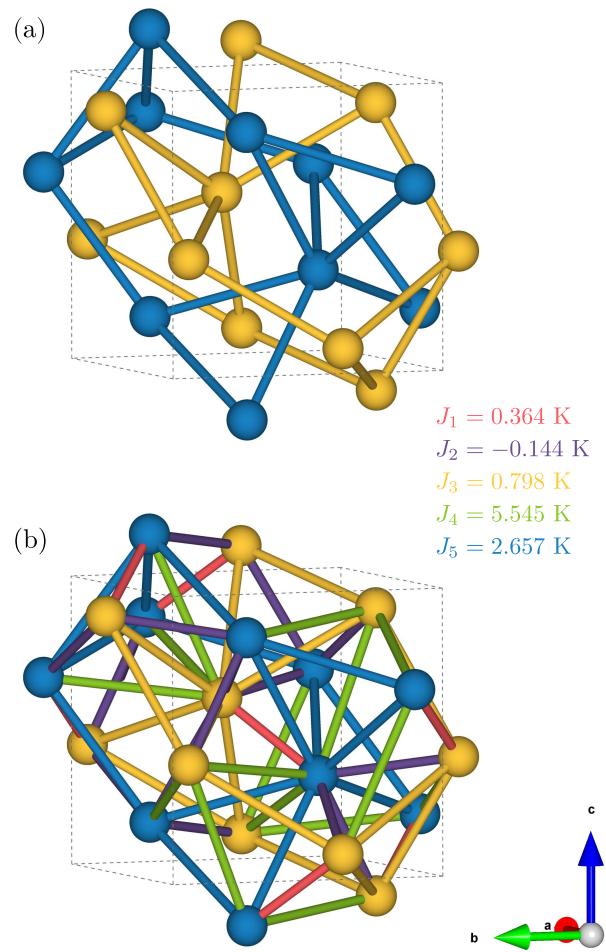}
    \caption{(a) The two trillium lattices: strong-TL strongly coupled by $J_5$ (blue) and weak-TL weakly coupled by $J_3$ (yellow). (b) The full double-trillium lattice and exchange interactions for K$_2$Ni$_2$(SO$_4$)$_3$ as derived in Ref.~\cite{Gonzalez24}.}
    \label{fig:latt}
\end{figure}

In Fig.~\ref{fig:latt} we show the magnetic structure of K$_2$Ni$_2$(SO$_4$)$_3$, where the Ni ions form two independent trillium lattices, as can be seen in the top panel in blue and yellow. The material can be described in terms of a cubic unit cell with 8 magnetic sites per unit cell, and the most relevant exchange interactions were derived with different methods, such as the density-functional theory energy mapping and self-consistent Gaussian approximation methods~\cite{Zivkovic21, Yao23, Gonzalez24}. Even though the different methods agree on the hierarchy of interactions, only one reproduces correctly the experimentally observed tripling of the magnetic unit cell~\cite{Gonzalez24}. These couplings are illustrated in Fig.~\ref{fig:latt}, where the two trillium lattices are formed by $J_3=0.798$~K (weak-TL in yellow) and $J_5=2.657$~K (strong-TL in blue), with $J_3$ considerably smaller than $J_5$. The largest coupling is $J_4 = 5.545$~K and connects the two trillium lattices (green bonds). Finally, $J_1=0.364$~K and $J_2=-0.144$~K (red and purple bonds) are much smaller than the rest and also represent connections between different trillium lattices, strong-TL and weak-TL. 

\subsection{Pulsed magnetic field}

The pulsed-field magnetization was measured at the Dresden High Magnetic Field Laboratory. We used single-crystal samples with a magnetic field along the $[11\bar{2}]$ direction. The magnetic field was produced by a solenoid energized by a 1.44 MJ capacitor bank module. The field rise time was 14 ms, with the total pulse duration about 100 ms. Magnetization was obtained by integrating the voltage induced in a compensated coil system surrounding the sample. Each measurement was corrected for background recorded under identical conditions without the sample~\cite{Skourski2010}. We kept the temperature of the sample stable by pumping on a $^3$He bath.

\subsection{Classical Monte Carlo}

The Heisenberg Hamiltonian of K$_2$Ni$_2$(SO$_4$)$_3$ in a magnetic field is given by
\begin{equation}\label{eq:ham}
    \mathcal{H} = \sum_{k=1}^5 \sum_{\langle ij \rangle_k} J_k \mathbf{S}_i \mathbf{S}_j - h \sum_i^N S_i^z,
\end{equation}
where $J_k$ are the exchange interactions from Fig.~\ref{fig:latt} and $\mathbf{S}_i$ are spin-1 operators. Because reliable quantum many-body methods for simulating this complex system are lacking, we instead employ cMC calculations. This approach is justified by the enhanced spin magnitude and the three-dimensionality, which are both expected to suppress quantum effects, and it is further supported a posteriori by the agreement with experimental results. In cMC we take $\mathbf{S}_i$ as classical 3-component vectors with $|\mathbf{S}_i|=1$. $S_i^z$ is the component in the $z$ direction subject to a magnetic field $h$. Since the model is spin-isotropic when $h=0$, the magnetic field can be applied in any arbitrary direction without changing the results. We consider cubic systems with periodic boundary conditions and $N=8L^3$ spins. To fit the magnetic unit cell at $h=0$, we take $L=3$, 6, 9, 12~\cite{Gonzalez24}.

Our cMC simulations are carried out starting from $T^*=2~J_4$ and cooling logarithmically down to $T^*=0.001~J_4$ with 200 temperature steps. After, we perform 10 equally spaced temperature steps to reach zero temperature. At each temperature, we perform between $3\times10^5$ and $1\times10^6$ Monte Carlo steps for the largest and smallest system sizes, respectively. For each Monte Carlo step, we perform $N$ Metropolis trials and $2N$ overrelaxation steps. The Metropolis trials are performed using the adaptive Gaussian step to guarantee a $50\%$ acceptance ratio~\cite{Alzate19}. Finally, we form 10 independent runs for each system size and magnetic field, and average the resulting measurements.

\section{Results}
\label{sec:res}

\subsection{Magnetization curve}

\begin{figure*}[t!]
    \centering
    \includegraphics[width=1.0\textwidth]{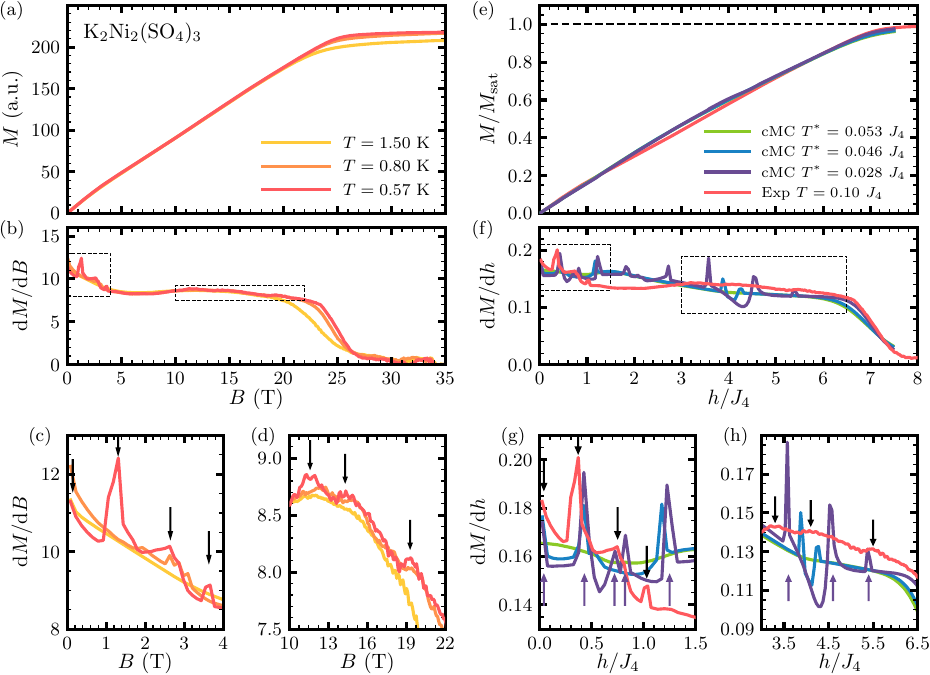}
    \caption{(a) Pulsed magnetic field measurement of the magnetization and (b) its derivative as a function of the magnetic field $B$ for the K$_2$Ni$_2$(SO$_4$)$_3$ langbeinite. (c,d) A zoom-in into the regions enclosed by the dashed black squares in panel (b). The arrows indicate four possible transitions in the low-field region and three in the intermediate-field region. Panels (e-h) show the same quantities for the $L=12$ cMC calculations at three different temperatures (green, blue, and purple), along with the experimental measurement (red). The purple arrows indicate the positions of the cMC peaks at the lowest temperature (purple curve). Note that the cMC temperatures have to be multiplied by $S(S+1)=2$ to be comparable with the experimental temperatures.}
    \label{fig:magcomp}
\end{figure*}

We show our high pulsed magnetic field data for K$_2$Ni$_2$(SO$_4$)$_3$ in Fig.~\ref{fig:magcomp}(a-d). The data for $T=1.5$~K, $0.8$~K, and $0.57$~K are presented in yellow, orange, and red, respectively. The system reaches the saturation for $B \geq 25$~T. The magnetization curve displays two distinct regimes. At low fields, the magnetization increases faster, as can be seen by the higher value of the derivative for fields $B \leq 4$~T [see Fig.~\ref{fig:magcomp}(b)]. From there on, and up to the saturation field, the magnetization grows almost linearly without any prominent features. However, the derivative $\mathrm{d}M/\mathrm{d}B$ evidences the appearance of several features at both low and intermediate field values as the temperature is lowered [see Fig.~\ref{fig:magcomp}(b-d)]. At low fields, there are several peaks which can be seen in the enlarged view on Fig.~\ref{fig:magcomp}(c) (see arrows). We note that they are not visible at temperatures above $T_c=1.1$~K, indicating they are associated with the ordered phase. Importantly, at intermediate fields [shown in Fig.~\ref{fig:magcomp}(d)] we find additional features which highlight field-induced phases and the potential reentrant behavior (see arrows).

To be able to directly compare the experimental results with cMC calculations, we extract the magnetization data at a given cMC temperature $T^*$ from different field runs $M_h(T^*)$ to build $M(h, T^*)$ and calculate its derivative. Additionally, to compare with our experimental measurements, we need to set them on an equal unit scale. We thus transform the experimental magnetic field $B$ in Tesla to dimensionless units as $B\to h/J_4 = \frac{g \mu_B}{k_B J_4} B$ with $g=2.34$~\cite{Zivkovic21}. On the other hand, the magnetization $M$ is normalized to be $M/M_\mathrm{sat}=1$ in the saturated region.

The cMC results for $L=12$ are shown in Fig.~\ref{fig:magcomp}(e-h), where the rescaled experimental data is shown in red and corresponds to the lowest measured temperature, $T=0.57$~K $=0.10~J_4$ [the same one as in Fig.~\ref{fig:magcomp}(a-d)]. We show three different cMC temperatures, $T^*=0.053~J_4$ above the zero-field finite-temperature phase transition at $T^*_c = 0.048(2)~J_4$ (green)~\cite{Gonzalez24}, and $T^*=0.046~J_4$ and $0.028~J_4$ below the phase transition (blue and purple). We note that a direct comparison of cMC and experimental temperatures requires multiplying the former by $S(S+1)=2$. We see in Fig.~\ref{fig:magcomp} that the cMC calculations agree quite well with the experimental results, even though there are deviations at intermediate fields around $h=3~J_4$. However, it is still surprising that the classical model reproduces so accurately the saturation field, which we find to be $h_c = 7.30(5)~J_4$ in cMC compared to the value we obtain for K$_2$Ni$_2$(SO$_4$)$_3$, that is about $h \sim 7~J_4$. This is one more piece of evidence in favor of the set of couplings we use here, already validated in Ref.~\cite{Gonzalez24}. 

As in the experimental data, the cMC results of the magnetization do not display any strong features, but many phase transitions are revealed when taking the field derivative. In Fig.~\ref{fig:magcomp}(f), we observe several peaks at different fields (see blue and purple curves). In the low-field region, shown in Fig.~\ref{fig:magcomp}(g), we see that cMC displays a very similar succession of phase transitions with five peaks in the range $0 \leq h/J_4 \leq 1.5$, denoted by the up-pointing purple arrows. These should be compared with the four down-pointing black arrows indicating the positions of the experimental peaks. Even though there is an extra peak in the cMC curve, there are two of them which are close together at about $h/J_4 \sim 0.8$, that could be associated with the third broad experimental peak. These phase transitions are visible in the specific heat phase diagram shown in Fig.~\ref{fig:cvs}, where the blue and purple lines and circles in the bottom panel correspond to the field windows and peaks appearing in Fig.~\ref{fig:magcomp}(g) and (h) (for the same blue and purple temperatures). Despite the theory-experiment agreement, it is hard to extract reliable conclusions about this succession of transitions. This model has an enlarged magnetic unit cell as ground state at $h=0$, consisting of $3^3 = 27$ structural unit cells, and small changes in the exchange parameters have been shown to lead to $4^3 = 64$ and $5^3 = 125$ enlargements~\cite{Gonzalez24}. The energies of these competing phases do not necessarily have the same field dependence, and thus small fields can produce many phase transitions (as we observe here). Since we are now restricting ourselves to $L=3n$ systems, many of these phases can be frustrated by the periodic boundary conditions, implying that the thermodynamic-limit phase diagram at low fields may differ slightly from the one we observe here. Thus, in what follows, we will only focus on the intermediate-field region, where our results are more reliable.

In the other zoom-in region in Fig.~\ref{fig:magcomp}(h), we observe two phase transitions whose separation grows when lowering the temperature from $T^*=0.046~J_4$ to $0.028~J_4$ (blue and purple curves). As the temperature is lowered (from blue to purple), a third peak appears at higher fields (see purple arrows), in agreement with the three experimental peaks observed. This peak comes from the splitting of the transition into two, as can be seen by the lines and dots in the bottom panel of Fig.~\ref{fig:cvs}. The third transition, which appears at the highest field, reaches the saturation field at zero temperature (see Fig.~\ref{fig:cvs}). Altogether, we observe a high level of correspondence between experimental results and cMC calculations. This supports our assumption that, despite the absence of quantum fluctuations in cMC, the classical model captures the essential magnetic properties of the material, thereby justifying the use of these simulations to interpret the experimental data further.

\subsection{Phase diagram}

\begin{figure}[t!]
    \centering
    \includegraphics[width=0.95\linewidth]{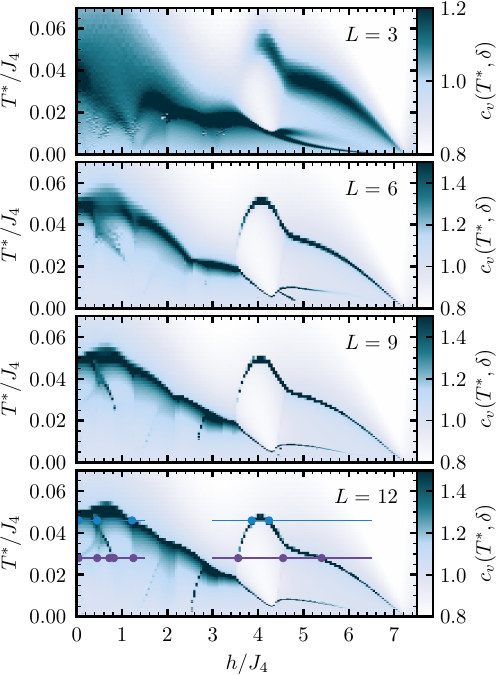}
    \caption{Specific heat $c_v$ as a function of temperature $T/J_4$ and magnetic field $h/J_4$ for four different system sizes $L=3$, 6, 9, and 12 (from top to bottom). The lines in the $L=12$ panel correspond to the temperatures shown in Fig.~\ref{fig:magcomp} in blue and purple, and span the field windows shown in panels (g) and (h). The dots over these lines correspond to the peaks observed in Fig.~\ref{fig:magcomp}.}
    \label{fig:cvs}
\end{figure}

To understand the origins of the features and transitions in the magnetization curve, we analyze the cMC phase diagram here. For this, we calculate the energy per site, $e$, and the specific heat per site, $c_v$, for fields ranging from $h=0$ to $7.5~J_4$ at $0.05~J_4$ intervals. We show in Fig.~\ref{fig:cvs} our results for the specific heat as a function of the temperature and magnetic field, for different system sizes. As mentioned above, we use $L=3n$ to fit the ground state in the $h=0$ and $T^*=0$ limit~\cite{Gonzalez24}. In this limit, $T_c = 0.048(2)~J_4$ and can be rescaled as $T_c^* = S(S+1)T_c = 0.096(4)~J_4$ to compare with the experimental value $T_c^\mathrm{exp} = 0.20~J_4$, which is only a factor 2 larger. The reasonable agreement can be explained by the $S=1$ value and the three-dimensionality of K$_2$Ni$_2$(SO$_4$)$_3$, both of which tend to suppress the effect of quantum fluctuations (although not completely). As the field increases, our calculations exhibit a complicated landscape of phase transitions and crossovers up to about $h=3.5~J_4$. 

These features at low fields are responsible for the many peaks observed in the derivative of the magnetization curve shown in Fig.~\ref{fig:magcomp}. In the bottom panel of Fig.~\ref{fig:cvs}, we have marked the regions (lines), temperature cuts (colors), and transitions (circles) corresponding to Fig.~\ref{fig:magcomp}(g) following the same color coding for the temperatures. On the other hand, the phase diagram above $h=3.5~J_4$ becomes quite simple and clean. The phase transition peak splits in two, with the upper branch showing a sharp increase in the critical temperature and forming a dome between $h=3.5~J_4$ and $4.5~J_4$. After this, the critical temperature decreases and splits into two again (see purple line in the bottom panel of Fig.~\ref{fig:cvs}). This splitting is the one that causes the evolution from two to three peaks when lowering the temperature in the field derivative of the magnetization [see blue and purple curves in Fig.~\ref{fig:magcomp}(h)]. The transition at the highest field finally vanishes at zero temperature when the system reaches the saturated ferromagnetic phase, while the remaining two form a dome around $h=4~J_4$. This part of the phase diagram resembles the phase diagram of the triangular lattice Heisenberg antiferromagnet in a magnetic field~\cite{Seabra11}, where the phases between the 3 intermediate-field transitions were identified as up-up-down and canted 2:1 phases. The transition between these two is a Berezinskii-Kosterlitz-Thouless transition, which would explain the faint peak in our specific heat calculations. Overall, our phase diagram exhibits small finite-size effects, indicating that most phase transitions are robust and likely to persist in the thermodynamic limit.

\subsection{Magnetization plateau}

\begin{figure}[t!]
    \centering
    \includegraphics[width=0.95\linewidth]{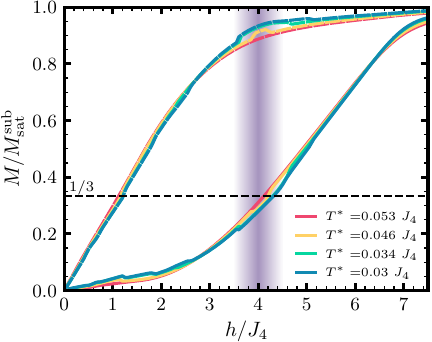}
    \caption{The cMC magnetization of the weak-TL and strong-TL sublattices (colored dashed and full lines, respectively), divided by the corresponding saturated value. The dashed black line indicates the value $M/M_\mathrm{sat}^\mathrm{sub} = 1/3$.}
    \label{fig:magind}
\end{figure}

The complicated arrangement of phase transitions below $h=3.5~J_4$ in Fig.~\ref{fig:cvs} can be associated with a subtle reordering of the magnetic structure. However, the more prominent feature, onto which we want to focus in the following, is the dome around $h=4~J_4$, where the critical temperature suddenly increases and decreases again. To understand the nature of this dome, it is instructive to separate the magnetization data by sublattice. This is shown in Fig.~\ref{fig:magind}, where the field dependence of the magnetization of the strong-TL and weak-TL trillium lattices is shown in full and dashed lines, respectively (see yellow and blue lattices in Fig.~\ref{fig:latt}). The two sublattices behave differently, due to the two trillium couplings being so distinct, $J_3\ll J_5$. The trillium sublattice weak-TL quickly aligns with the magnetic field (dashed lines) and reaches its saturation point at about $h=4~J_4$, which is where the dome in the specific heat peaks. On the other hand, the trillium sublattice strong-TL has a much slower increase in the magnetization and only has $1/3$ of the saturated value at $h=4~J_4$ (see black dashed line). The shaded region in Fig.~\ref{fig:magind} indicates that, as the temperature is lowered, weak-TL tends towards saturation, while strong-TL decreases its magnetization. 

\begin{figure}[t!]
    \centering
    \includegraphics[width=0.95\linewidth]{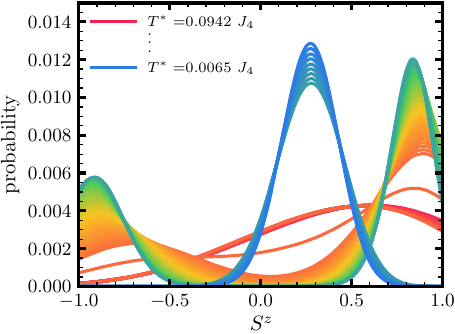}
    \caption{Histograms of the $S^z$ distribution of spins in the strong-TL trillium sublattice as a function of the temperature for a magnetic field crossing the dome, $h=4~J_4$. The highest temperature is displayed in red, and the lowest one in blue.}
    \label{fig:histo}
\end{figure}

When the weak-TL reaches the saturation field (beyond $h=4~J_4$), it acts as an effective field over the strong-TL, allowing us to write an effective Hamiltonian,
\begin{equation}\label{eq:heff}
    \mathcal{H}_\mathrm{eff} = J_5 \sum_{\langle ij\rangle_5} \mathbf{S}_i \mathbf{S}_j - \left(h -  \sum_{k}^4 n_k J_k \right) \sum_{i\in J_5}^{N/2} \mathbf{S}_i^z
\end{equation}
where the Heisenberg interaction only accounts for the $J_5$ trillium lattice and $n_k$ is the number of neighbors that a site in the $J_5$ trillium lattice has from the $J_3$ trillium lattice. Note that $n_3 = 0$ since $J_3$ does not connect to spins in the $J_5$ sublattice. On the other hand, $n_1=1$ and $n_2=n_4=3$ (see Fig.~\ref{fig:latt}). Knowing the saturation field $h_c=7.3~J_4$ of the Hamiltonian in Eq.~(\ref{eq:ham}) from cMC simulations, we can use Eq.~(\ref{eq:heff}) to predict the saturation field $h_c^\mathrm{trill}$ of the trillium lattice Heisenberg model with nearest neighbor coupling $J$. Specifically, we obtain
\begin{equation}
    \frac{h_c^\mathrm{trill}}{J} = \frac{1}{J_5} \left( h_c - \sum_{k}^4 n_k J_k \right) = 9.00.
\end{equation}
Furthermore, using the same equation and our observation that the phase diagram of K$_2$Ni$_2$(SO$_4$)$_3$ has a dome-like phase transition at $h=4~J_4$, we can infer that the trillium model has a similar feature at $h^\mathrm{trill}=2~J$ (we will confirm the accuracy of these predictions in Sec.~\ref{sec:plateau_order}). Therefore, the dome in the phase diagram is a signature of the trillium lattice in a magnetic field. As such, it is likely to be present in many members of the langbeinite family, and not only in K$_2$Ni$_2$(SO$_4$)$_3$, given that the two trillium lattices have different couplings (which is expected already from the structure). 

\begin{figure}[t!]
    \centering
    \includegraphics[width=0.95\linewidth]{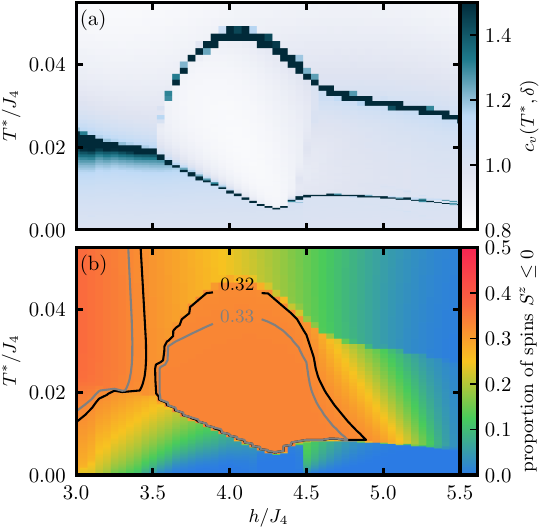}
    \caption{(a) Specific heat and (b) integral of the histograms of the $S^z$ distribution of spins in the strong-TL trillium sublattice as a function of the temperature and the magnetic field. The contour lines indicate the region where about $1/3$ of the spins point against the magnetic field.}
    \label{fig:histoint}
\end{figure}

To study the type of symmetry breaking and the phase within the dome, we need information from the spin configuration on the $J_5$ sublattice, since we know the $J_3$ sublattice is completely polarized. We do this by calculating the histogram of the distribution of $S^z$ for all spins in the $J_5$ sublattice at all temperatures. The results obtained for $h=4~J_4$ are shown in Fig.~\ref{fig:histo} for temperatures between $T^*=0.0942~J_4$ and $0.0065~J_4$, going inside and outside the dome. At high temperatures, above the phase transition (reddish colors), the spins point mostly \textit{up} with a wide peak around $\langle S^z\rangle \sim 0.5$. As the temperature is lowered and the system crosses the phase transition, there is a splitting of the distribution which becomes bimodal, with peaks close to $S^z=-1$ and $+1$. Furthermore, the height of the high-$S^z$ peaks is about double the height of the low-$S^z$ peak, indicating that $2/3$ of the spins point \textit{up} while $1/3$ points \textit{down} (opposing the magnetic field tendency). This is a clear indication that a $1/3$ magnetization plateau-like phase forms inside the dome on the strong-TL trillium subsystem, which can be translated into a $2/3$ magnetization plateau-like phase on the whole system (5 spins \textit{up} and 1 spin \textit{down}). This phase breaks the discrete translation symmetry of the lattice. We call this phase plateau-like because the magnetization plateaus are purely quantum phases preserved by a gap at $T=0$. In the classical case, plateaus do not exist at $T=0$, but appear as domes or bubbles at higher temperatures, as it happens, for example, on the triangular lattice~\cite{Seabra11}. Further lowering the temperature induces a second phase transition, as observed in Fig.~\ref{fig:cvs}, where the spins become all equivalent again, indicating that the discrete broken symmetry is recovered. This is reflected by the single peak in Fig.~\ref{fig:histo} for the lowest temperatures (blue colors).

Another way to see the existence of the plateau phase is to calculate the integral of the $S^z$ distribution from $S^z=-1$ to 0, to obtain the fraction of spins pointing against the magnetic field $h$. In Fig.~\ref{fig:histoint} we show this fraction for the $J_5$ trillium sublattice in the dome region of the phase diagram. Comparing with the phase diagram extracted from specific heat in the upper panel, we see that the region inside the dome contains about $1/3$ of the spins opposing the magnetic field. This quantity is about $0.5$ when the temperature is high, and the spins are completely disordered, or at low fields where there is no preferred $S^z$ direction. On the other hand, as the field increases and the temperature decreases, the integral approaches zero, since all spins align with the magnetic field.

\subsection{Magnetic order in the plateau phase}
\label{sec:plateau_order}

\begin{figure}[t!]
    \centering
    \includegraphics[width=0.95\linewidth]{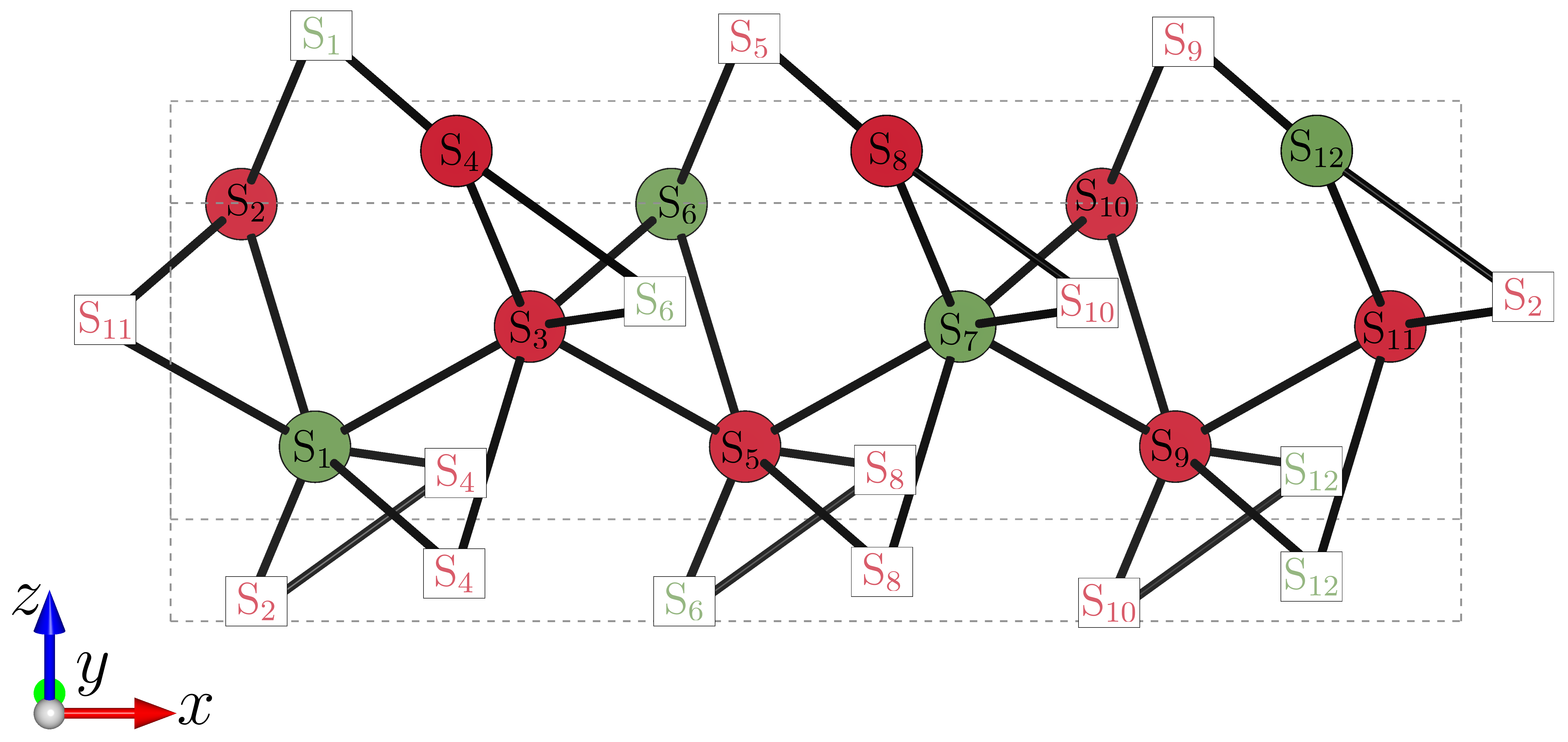}
    \caption{Magnetic unit cell of the up-up-down phase on the trillium lattice, with down spins illustrated in green and up spins in red. The 12 sites in the unit cell are shown by circles, and correspond to the 4 sites $(S_1,S_2,S_3,S_4)$ in the structural unit cell plus translations in $a$ $(S_5,S_6,S_7,S_8)$ and $2a$ $(S_9,S_{10},S_{11},S_{12})$, with $a$ being the size of the unit cell. The equivalent sites outside the unit cell (periodic boundary conditions or translation invariance) are shown in rectangles.}
    \label{fig:uud}
\end{figure}

The plateau phase implies an enlargement of the magnetic unit cell compared to the structural unit cell, because the four spins in the trillium unit cell cannot be split into two up and one down. We find that the up-up-down phase on the strong-TL trillium sublattice forms such that there are two up spins and one down spin in every triangle, as shown in Fig.~\ref{fig:uud} where up and down spins are indicated with red and green colors, respectively. This implies that the down spins on the $J_5$ trillium sublattice are connected to 6 up spins, and the up spins are connected to 3 up and 3 down spins, such that they have a zero effective field. This scenario is the same as the $1/3$ magnetization plateau on the triangular lattice. To fit this phase in the trillium lattice with periodic boundary conditions, at least along one cubic direction the system size has to be a multiple of 3, $L_i = 3n$ for $i=x$, $y$, or $z$; and thus the minimal magnetic unit cell is the $(L_x, L_y, L_z) = (3,1,1)$ cluster with 12 spins shown in Fig.~\ref{fig:uud}. There, we observe each spin in the structural unit cell appears pointing \textit{down} (green) once every three unit cells (for example, $\downarrow S_1$, $\uparrow S_5$, $\uparrow S_9$).  As shown in Ref.~\cite{Gonzalez26}, the up-up-down phase on the trillium lattice belongs to the ground-state manifold of the Ising or Heisenberg models on the trillium lattice. Even though, in principle, quantum fluctuations may cause an overlap of many such configurations, it was shown in Ref.~\cite{Gonzalez26} that there is only a very small number of degenerate configurations with $1/3$ magnetization in the Ising model. For example, in a lattice of size $(L_x,L_y,L_z) = (3,2,2)$ there are only 3 different spin configurations with $M=1/3$. While this might indicate a stability of the ordered state in Fig.~\ref{fig:uud} under quantum fluctuations, its precise quantum properties remain unclear at this point.

\begin{figure}[t!]
    \centering
    \includegraphics[width=0.95\linewidth]{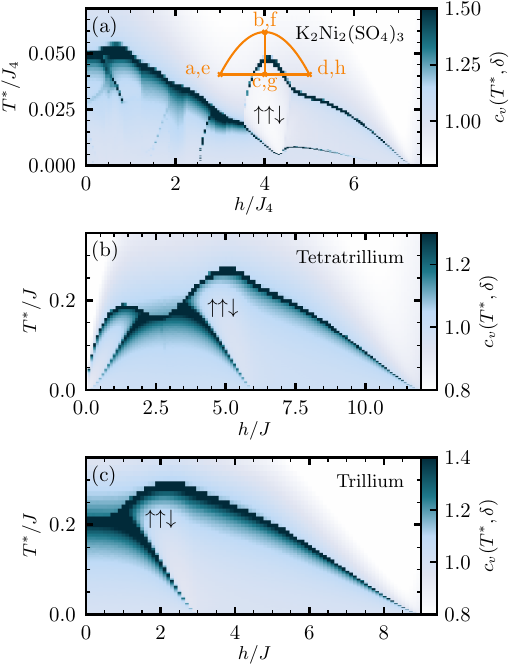}
    \caption{(a) Specific heat of the model corresponding to K$_2$Ni$_2$(SO$_4$)$_3$, (b) to the tetratrillium and (c) trillium lattices. The $\uparrow \uparrow \downarrow$ indicates the region of the plateau phase. Note that for K$_2$Ni$_2$(SO$_4$)$_3$ and the tetratrillium lattices, this implies that the remaining trillium lattice is fully polarized, producing a total $2/3$ magnetization plateau.}
    \label{fig:allcvs}
\end{figure}

\begin{figure*}[t!]
    \centering
    \includegraphics[width=0.99\linewidth]{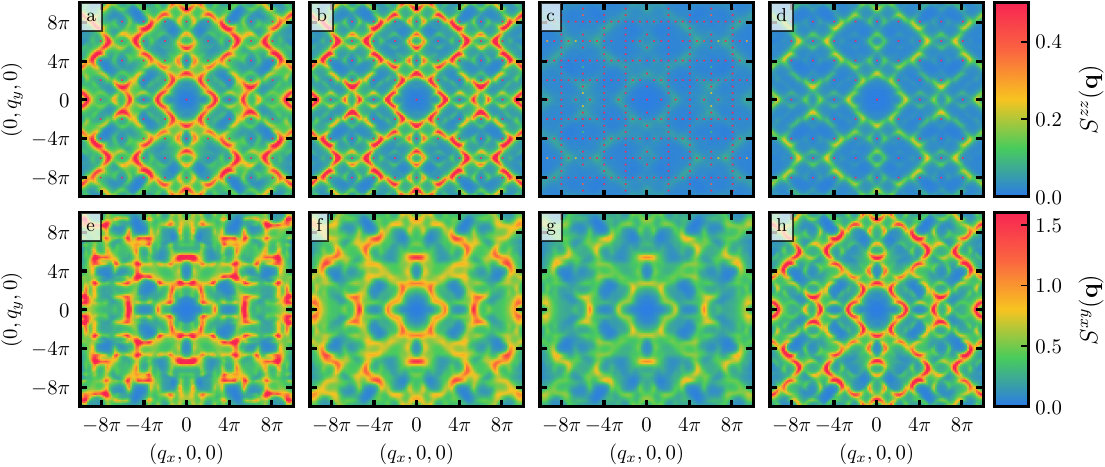}
    \caption{Spin structure factor obtained with cMC in the $q_x$-$q_y$ plane at $q_z=0$. The field and temperature values of each panel are denoted by the orange lines and crosses in the top panel of Fig.~\ref{fig:allcvs}. The top row shows the longitudinal spin structure factor $S^{zz}(\mathbf{q})$, where the Bragg peaks have a much stronger signal than the diffuse patterns and therefore appear as red dots. The bottom row shows the transverse spin structure factor $S^{xy}(\mathbf{q})$, perpendicular to the field. The color scale is consistent in each row.}
    \label{fig:ssf}
\end{figure*}

\subsection{Ubiquity of the plateau phase}

Here, we show the ubiquity of the plateau-like phase by studying the corresponding models on the trillium and tetratrillium lattices. Here, the trillium lattice model is obtained by setting $J_5=1$ and $J_1=J_2=J_3=J_4=0$ while the tetratrillium lattice model has $J_4=J_5=1$ and $J_1=J_2=J_3=0$. The results for the specific heat phase diagrams are shown in Fig.~\ref{fig:allcvs}, where the plateau-like phases are marked by the $\uparrow \uparrow \downarrow$ symbol. For the trillium case, the saturation field matches our prediction above, and we find $h_c=9~J$. Also, we find a $1/3$ plateau-like phase at around $h=2~J$, which extends down to a single point at zero temperature, at about $h=3~J$. On the other hand, the tetratrillium lattice has a larger saturation field of $h_c=12~J$. In this case, we also find a $ 2/3$-plateau-like phase. This corresponds, as in K$_2$Ni$_2$(SO$_4$)$_3$, to a fully polarized trillium sublattice while the other sublattice forms a 2-up-1-down type of order. Also, in this case, the phase appears to extend to zero temperature, where it collapses to a single point at $h=6~J$.

\subsection{Reentrant behavior}

Due to the dome shape of the up-up-down phase in the $T$-$h$-phase diagram, the system experiences a reentrant behavior when the magnetic field increases at constant temperature. This is shown by the horizontal straight orange line in Fig.~\ref{fig:allcvs}, which crosses the up-up-down phase. The phases at points (a,e) and (d,h) are the same because they can be continuously connected by the curved orange line that goes through (b,f), above the phase transition. To see this explicitly, we calculate the longitudinal and transverse spin structure factors $S^{zz}(\mathbf{q})$ and $S^{xy}(\mathbf{q})$, respectively, which are defined by
\begin{align}
S^{zz}(\mathbf{q})&=\frac{1}{N}\sum_{i,j}\ \langle S_i^z S_j^z\rangle \ e^{i\mathbf{q}(\mathbf{r}_i-\mathbf{r}_j)},\\
S^{xy}(\mathbf{q})&=\frac{1}{N}\sum_{i,j}\ \langle S_i^x S_j^x+S_i^yS_j^y\rangle \ e^{i\mathbf{q}(\mathbf{r}_i-\mathbf{r}_j)}.
\end{align}
Here, $N$ is the total number of sites and $\mathbf{r}_i$ is the position of site $i$. The results are shown in Fig.~\ref{fig:ssf}, where the upper row corresponds to $S^{zz}(\mathbf{q})$ and the lower to $S^{xy}(\mathbf{q})$. The letters assigned to each panel represent the temperatures and fields marked by the orange crosses in Fig.~\ref{fig:allcvs}. Focusing first on the $zz$ component, we can see that the spin structure factor along the (a,b,d) path evolves continuously: the Bragg peaks remain at the same positions, and the background signal gets weaker as the spins align more and more with the magnetic field. We also observe that the patterns are mostly formed by half-moons that surround the Bragg peaks. On the other hand, along the path (a,c,d), we can see that a new set of Bragg peaks appears at point (c) inside the up-up-down phase. Because several independent runs are averaged to produce this plot, the additional Bragg peaks appear in both $q_x$ and $q_y$ directions (also at non-zero $q_z$, not visible here). However, each independent run only contains additional peaks along one direction because, as we showed in Fig.~\ref{fig:uud}, the up-up-down phase has an enlarged magnetic unit cell in only one Cartesian direction. Another interesting feature is that the background signal is lower inside the up-up-down phase, implying that fluctuations are smaller. Finally, we can observe a similar behavior in the transverse spin structure factor (lower row), where the intensity of the signal is lower inside the up-up-down phase [see panel (g)] and becomes stronger once the system exits it [see panel (h)]. We can also conclude that there is no ordering tendency along the transverse direction, as no sharp peaks appear, and the signal is similar to that of the longitudinal direction, but without Bragg peaks. 

\section{Conclusions}
\label{sec:conc}

We studied the magnetization process of the double-trillium langbeinite K$_2$Ni$_2$(SO$_4$)$_3$ with $S=1$, using experimental and theoretical methods. On the experimental side, we performed high pulsed-field measurements on single-crystal samples to reach the ferromagnetically saturated region at about $B\sim 25$~T. Using the derivative of the magnetization curve, we uncovered the presence of several features at low fields and intermediate fields. To explain these observations, we studied the classical Hamiltonian associated with K$_2$Ni$_2$(SO$_4$)$_3$, which has already proven accurate in describing the ground-state ordering and static spin structure factor. Here, we found that the Hamiltonian also provides an accurate estimate of the saturation field.

The classical calculations also find a landscape of several phase transitions at low magnetic fields. We attributed these to the many competing phases with enlarged magnetic unit cells, which are challenging to resolve on finite lattices. We deferred an analysis of these features to future work and instead focused on the transitions at intermediate fields. We found two transitions with cMC that correspond to the existence of a dome in the temperature-field phase diagram. Interestingly, inside the dome, one-sixth of the spins oppose the magnetic field, forming a classical precursor of a $2/3$ magnetization plateau. In this phase, all spins in the weakly coupled $J_3$ trillium lattice point in the direction of the field. Meanwhile, in the strongly coupled $J_5$ trillium lattice, two-thirds of the spins align with the field, and one-third oppose it, resembling the up-up-down phase on the triangular lattice.

We studied this phase in detail, finding an enlarged magnetic unit cell with 12 spins (three times the structural unit cell). Surprisingly, the up-up-down state fulfills the ground-state conditions of the Ising model on the trillium lattice, with two spins pointing \textit{up} and one pointing \textit{down} in each triangle. This causes every \textit{down} spin to be surrounded by 6 \textit{up} spins, and every \textit{up} spin to be surrounded by 3 \textit{up} and 3 \textit{down} neighbors. Also, we revealed that this up-up-down phase is very common in these types of lattices and is present both on the single trillium and tetratrillium lattices. This implies that this phase might be present in many members of the langbeinite family, consisting of coupled trillium lattices.

Finally, we studied the reentrant behavior produced by the existence of the up-up-down dome. The dome is responsible for two phase transitions in and out of the plateau phase. Since the magnetic unit cell is enlarged inside the dome, the spin structure factor exhibits a distinctive pattern of additional Bragg peaks. On the other hand, the phase at lower and higher fields presents the same pattern of Bragg peaks associated with the ferromagnetic ordering due to the magnetic field, with the same structure of fluctuations in the perpendicular component and in the background of the longitudinal one. 

Overall, these results raise several important questions for both experiment and theory. Experimentally, it remains to be seen whether trillium and double-trillium compounds exhibit the magnetization plateaus and the associated up–up–down order predicted by our analysis. On the theoretical side, the stability of these magnetization plateaus in the quantum $S=1/2$ and $S=1$ cases remains an open problem.

\section*{Acknowledgments}

M.G.G. gratefully acknowledges the access to the Marvin cluster of the University of Bonn and the HPC Service of FUB-IT, Freie Universit\"at Berlin, for computing time. We acknowledge the support of HLD at HZDR, a member of the European Magnetic Field Laboratory (EMFL).

\bibliography{biblio}

\end{document}